\documentclass[12pt,tightenlines,eqsecnum,floats,aps,amsmath,amssymb,nofootinbib,prd,superscriptaddress]{revtex4}
\usepackage{setspace}
\usepackage{amsmath,amssymb,amsfonts,amsthm,mathrsfs}
\usepackage{graphicx}
\usepackage{enumerate} 
\usepackage{comment}
\usepackage{slashed}
\usepackage[pdftex]{hyperref}
\usepackage[normalem]{ulem}
\def\be{\begin{equation}}
\def\ee{\end{equation}}
\def\ba{\begin{eqnarray}}
\def\ea{\end{eqnarray}}
\def\bi{\begin{itemize}}
\def\ei{\end{itemize}}

\def\reals{\mathbb{R}}

\def\xh{\hat{x}}
\def\qh{\hat{q}}
\def\ph{\hat{p}}
\def\w{\omega}
\def\dpp{\widetilde{d p}\,}
\def\K{\mathbb{K}}
\def\Qq{Q(\qh)}
\def\Qqs{Q^{\rm soft}(\qh)}
\def\Qqh{Q^{\rm hard}(\qh)}
\def\ss{\sigma^{\rm soft}}
\def\sh{\sigma^{\rm hard}}
\def\phit{\tilde{\phi}}
\def\Qs{Q^{\rm soft}}
\def\Qh{Q^{\rm hard}}
\def\t{\tau}
\def\H{\mathcal{H}}
\def\I{\mathcal{I}}
\def\phih{\varphi_{\H}}
\def\D{\mathcal{D}}
\def\G{\mathcal{G}}
\def\vol{\text{Vol}}
\def\Lh{\Lambda_{\H}}

\begin{document}

\title{Asymptotic charges from soft scalars in even dimensions}

\author{Miguel Campiglia}
\author{Leonardo Coito}
\affiliation{Instituto de F\'isica, Facultad de Ciencias,  Montevideo 11400, Uruguay}

\begin{abstract}
We study asymptotic charges associated to  a  spin-zero analogue of Weinberg's soft photon and graviton theorems in even dimensions. Simple spacetime expressions for the charges are given, but unlike gravity or electrodynamics, the symmetry interpretation for the charges remains elusive. This work is a higher dimensional extension of the four dimensional case studied in \cite{ccm}. 

  \end{abstract}

\maketitle
\tableofcontents

\section{Introduction}
Soft factorization theorems in scattering amplitudes can be understood as  conservation laws of asymptotic charges, see \cite{lecstrom} for a recent review. When the soft particle has spin $1$ or $2$, the asymptotic charges are associated to gauge symmetries that are non-trivial at infinity \cite{lecstrom}. For spin zero soft particles one can still construct conserved asymptotic charges \cite{ccm,hamada}, but their interpretation in terms of symmetries remains incomplete.  The factorization studied in \cite{ccm} is a `leading' soft theorem (a spin-zero analogue of Weinberg's soft theorem \cite{weinberg}), whereas \cite{hamada} deals with factorization at subleading order in the soft scalar energy.  

In the hope to improve the understanding of spin-zero asymptotic charges, here we extend the results of \cite{ccm} to higher even dimensions. One of the advantages is that  we can  study massless $\varphi^3$ theory (which  does not have a good null-infinity description in dimension four, see section \ref{scrisec}). This is perhaps the simplest context where to explore the question of asymptotic symmetries associated to soft scalars. %\footnote{Massless $\varphi^3$ theory was excluded in the treatment of \cite{ccm} since its $1/r$ expansion is ill-defined in dimension four. See footnote \ref{fnd4}.}  
Furthermore, it includes (upon incorporating internal indices) the case of the bi-adjoint scalar theory \cite{biadjoint}, whose asymptotic symmetries one hopes \cite{ccm} could be related to those in Yang-Mills through a `double-copy' \cite{white}.  As in \cite{ccm} our discussion will be restricted to tree-level amplitudes. The study of loop corrections would require  to work in a theory where masslessness of the scalar is protected (as for instance the model studied in \cite{hamada}). 

%It is also relevant  theory where one can study  An understanding of asymptotic charges in $\varphi^3$ theory could help elucidate the underlaying symmetry. It is also needed if one wishes  to explore  asymptotic symmetries from a `double-copy' \cite{white}  perspective. It may also be relevant for recent approaches to flat-space holography \cite{}.

%As in the higher dimensional cases of QED \cite{} and gravity \cite{}, the relation between 

The organization of this paper is as follows. In section \ref{sec2} we recall the charge associated to the soft-scalar theorem in \cite{ccm}. In section \ref{sec3} we interpret this charge in terms of asymptotic fields at null infinity for massless $\varphi^3$ theory. In section \ref{sec4} we extend the analysis to include  massive fields $\psi$ coupled to the massless scalar. In section \ref{sec5} we discuss smeared version of the charges and give a simple expression for a spacetime current. Some of the calculations are given in appendices. 

Our conventions are as follows. We take spacetime dimension to be $D=2m+2$ with $m=2,3,\ldots$. We use mostly plus signature spacetime metric. For simplicity in our discussion we focus on \emph{future} null and time infinities; a parallel discussion applies to the past infinities. We also omit a discussion of fields at spacelike infinity. This shares many  features with the study of fields at time-infinity, except that differential equations become hyperbolic. In particular, it allows one to understand the conservation of asymptotic charges from a dynamical perspective \cite{eyhe,ccm}.

\section{Soft theorem charge} \label{sec2}

We consider a theory of a self-interacting scalar field $\varphi$ coupled to a massive scalar $\psi$,\footnote{We could also consider  $\psi$'s with nonzero spin \cite{ccm} or  with zero mass (in which case the treatment will be similar to that of $\varphi^3$ theory). The main reason to introduce $\psi$ is to discuss how to deal with massive fields.}  
\be
L= -\frac{1}{2} (\partial \varphi)^2 - \frac{1}{2} (\partial \psi)^2 - \frac{1}{2}\mu^2 \psi^2+ \frac{g}{3!} \varphi^3 + \frac{g'}{2!}  \psi^2 \varphi   . \label{Lint}
\ee
At tree level, one can obtain a `soft theorem' that relates an $(n+1)$-particle amplitude  with an $n$-particle amplitude, where the $(n+1)-$th particle is a soft $\varphi$. This soft theorem can be written as matrix elements of the identity \cite{lecstrom}
\be
\Qq S = S \Qq, \label{QSSQ}
\ee
where $S$ is the S-Matrix, $\qh$ the direction of the soft $\varphi$-particle and \cite{ccm} % the soft theorem identity can be written as a conservation law  of the charge
\be
\Qq:= \lim_{\w \to 0}\frac{\w}{2}(a(\w \qh) +a^\dagger(\w \qh)) -\frac{g}{2} \int \dpp \frac{a^\dagger(\vec{p})  a(\vec{p})}{p \cdot q} -\frac{g'}{2} \int \dpp \frac{b^\dagger(\vec{p})  b(\vec{p})}{p \cdot q}.
\label{Qq}
\ee
Here $a$ and $b$ are the Fock operators associated with the fields $\varphi$ and $\psi$ respectively, $\dpp$ the standard Lorentz invariant measure and 
\be
q^\mu=(1,\qh) \label{qmu}
\ee
a null, future pointing four-momentum associated to $\qh$.

\section{Charges in massless $\varphi^3$ theory} \label{sec3}

In this section we set $g'=0$ in (\ref{Lint}) and restrict attention to  massless $\varphi^3$ theory. We remind the reader that we  work in even spacetime dimension grater than four: $D=2m+2$ with $m=2,3,\ldots$.

\subsection{Field expansion near null infinity} \label{scrisec}
We assume the standard $1/r$ expansion off null infinity,
\be
\varphi(r,u,\xh)= r^{-m}\sum_{n=0}^{m-1} r^{-n}\phi_n(u,\xh) + O(r^{-2m}). \label{phiexp}
\ee
As we shall see, the leading term $\phi_0$ represents the `free data' in terms of which  the remaining $\phi_n$'s can be expressed by  recursively solving the field equations.  The field equation for $\varphi$ is: 
\be
\square \varphi + \frac{g}{2} \varphi^2 =0 \label{fieldeq}
\ee
where, in retarded coordinates, the wave operator takes the form
\be
\square \varphi =  r^{-m}\left[ - 2 \partial_u \partial_r (r^m \varphi)  + \partial^2_r (r^m \varphi)\right]+ r^{-2}[D^2 -m(m-1)]\varphi
\ee
with $D^2$ the Laplacian on the $2m$-dimensional sphere.  Plugging the expansion (\ref{phiexp}) in (\ref{fieldeq}) one finds:
\be
2n \partial_u \phi_n = -[D^2-(m-n)(m+n-1)] \phi_{n-1} \quad \text{for} \quad n=1,\ldots,m-2. \label{phineq}
\ee
For higher $n$ one gets $O(g)$ contributions from the second term in (\ref{fieldeq}). For $n=m-1$ this gives:\footnote{In four space-time dimensions  equation (\ref{phimeq}) implies  $\phi_0=0$ and so the expansion is not consistent with the field equations. Relaxing the fall-offs  (\ref{phiexp}) to include logarithmic terms (such as  $\log r/r$)  does not  fix things. \label{fnd4}}
\be
2(m-1)\partial_u \phi_{m-1} = -[D^2-2(m-1)] \phi_{m-2} - \frac{g}{2} \phi_0^2. \label{phimeq}
\ee
%with the coefficient of $\phi_{m-2}$ as in (\ref{phineq}).  
 We will see that  $\phi_{m-1}$ plays a role analogous to `Bondi mass-aspect' in gravity or `charge aspect' in Maxwell theory. In particular, its $u \to -\infty$ asymptotic value defines an angular charge density $\sigma(\xh)$,
\be
\sigma(\xh) :=  \phi_{m-1}(u=-\infty,\xh) \label{sigma}
\ee
which, as we shall see, is conserved and encodes the same information as the charge $\Qq$ defined in Eq. (\ref{Qq}).

%We will see that the charge $\Qq$ (\ref{Qq}) can be written in terms of the $u \to -\infty$ asymptotic value of $\phi_{m-1}$:
%\be
  %\phi_{m-1}(u=-\infty,\qh) \equiv - \int_{-\infty}^\infty du \partial_u \phi_{m-1}(u,\qh)  + \phi_{m-1}(u=\infty,\qh). \label{Qqphi}
%\ee

Before moving on, we need to specify the $u \to \pm \infty$  fall-offs implicitly assumed in (\ref{sigma}) and consistent with (\ref{phimeq}).  %Eq. (\ref{sigma}) implies certain $u \to -\infty$  fall-offs that we now specify.  %In order to  make sense of (\ref{sigma}) and later calculations,  we first need to specify $u \to \pm \infty$ fall-offs for $\phi_n$. 
We require:
\be
\begin{array}{ccl}
\phi_n(u,\xh)  & = &  O(|u|^{-(m-1-n+\epsilon)}) \quad \text{for} \quad n=0,\ldots,m-2  \label{fallu} \\
 \phi_{m-1}(u,\xh) & = &O(1)+O(|u|^{-\epsilon})
\end{array}
 \ee
 as $u \to \pm \infty$ for some $\epsilon>0$.  These conditions may  be thought of as  real-space version of the condition that $\w a(\w,\qh)$ has a well-defined $\w\to 0$ limit (see Eq. (\ref{Qq}) and the  next subsection).

With these conditions we can now use Eqns. (\ref{phineq}), (\ref{phimeq}) to write $\partial_u \phi_{m-1}$ in terms of the free-data $\phi_0$. Introducing the notation:
\be
\Delta_n := D^2-(m-n)(m+n-1),
\ee
one finds\footnote{See \cite{mirbabayi,stromhdqed} for analogous expressions in QED.}
\be
\partial_u \phi_{m-1}= \frac{1}{(-2)^{m-1}(m-1)!} \prod_{n=1}^{m-1} \Delta_n  \left[\int^u_{-\infty} du \right]^{m-2} \phi_0 - \frac{g}{4(m-1)}\phi_0^2 \label{duphim}
\ee
where $\left[\int^u_{-\infty} du \right]^{m-2} \phi_0$ denotes the $(m-2)$-th primitive of $\phi_0(u)$.

\subsection{Relation between $\Qq$ and $\sigma(\xh)$}
In this section we use the   sphere-differential operator appearing in (\ref{duphim})  to relate the charge $\Qq$ (\ref{Qq}) with  $\sigma(\xh)$ (\ref{sigma}). %Conservation of the former follows from the soft theorem, whereas conservation of the latter may be established by studying the asymptotic field equations at spatial infinity as in \cite{}.
Define the $2(m-1)-$th differential operator:
\be
\K := -\frac{1}{(-4 \pi)^m (m-1)!} \prod_{n=1}^{m-1} \Delta_n  \label{K}
\ee
In appendix \ref{appK} it is shown that $\K$ is invertible, with inverse given by the  Green's function
\be
K^{-1}(\qh,\xh) := \frac{2(m-1)}{1-\qh \cdot \xh} \; , \quad  \quad \K \, K^{-1}(\qh,\xh) = \delta^{(2m)}(\qh,\xh). \label{Kinv}
\ee
We show below that $\K$ maps  $Q(\xh)$ into $\sigma(\xh)$:
\be
 \sigma(\xh) =  \K \,  Q(\xh)  ,  \label{sigQ}
\ee
with inverse relation 
\be
Q(\qh)   = \int_{S^{2m}} d^{2m} \xh \, K^{-1}(\qh,\xh) \sigma(\xh). \label{Qsig}
\ee
To establish these relations we compare the   `soft' and `hard'  parts of $\Qq$ and $\sigma(\xh)$  separately. The soft and hard parts of   $\Qq$ are:
\ba
\Qqs & = & \lim_{\w \to 0^+}\frac{\w}{2}(a(\w \qh) +a^\dagger(\w \qh)) \label{Qs} \\
\Qqh & = &  -\frac{g}{4 (2 \pi)^{2m}} \int_{S^{2m}} d^{2m} \ph \int_{0}^{\infty} \frac{d E}{2 \pi} \,   \frac{E^{2m-2} a^\dagger(E \ph)  a(E \ph)}{-1 + \qh \cdot \ph} , \label{Qh}
\ea
where in writing $\Qqh$ we expressed  the momentum integral in spherical coordinates $\vec{p}= E \ph$. For $\sigma(\xh)$ we start with  definition (\ref{sigma}) to obtain:
\be
\sigma(\xh)=- \int_{-\infty}^\infty du \partial_u \phi_{m-1}(u,\xh)  + \phi_{m-1}(u=\infty,\xh). \label{sigma2}
\ee
As we shall see in the next section, the term  $\phi_{m-1}(u=\infty,\xh)$ accounts for  massive particles  but is zero in the massless theory.  The term  $\partial_u \phi_{m-1}(u,\xh)$ can be expressed in terms of the free-data $\phi_0$ by means of  Eq. (\ref{duphim}), yielding  `soft' (linear in $\phi_0$) and `hard' (quadratic in $\phi_0$) pieces,
\be
\sigma(\xh) = \ss(\xh) + \sh(\xh).
\ee
To compare with $\Qq$ we express $\sigma^{\rm soft/hard}$ in terms of the Fourier transform of $\phi_0$. Defining
\be
\phit_n(E,\xh)= \int_{-\infty}^{\infty} \phi_n(u,\xh) e^{i E u} du,
\ee
one finds
\ba
\ss(\xh) & = & - \lim_{E \to 0}   \frac{1}{(-i E)^{m-2}(-2)^{m-1}(m-1)!} \prod_{n=1}^{m-1} \Delta_n \phit_0(E,\xh) \label{sigmas}\\
\sh(\xh)& = &  \frac{g}{2(m-1)}\int_{0}^{\infty} \frac{dE}{2 \pi} \phit_0(E,\xh) \phit_0(-E,\xh).\label{sigmah}
\ea
We finally express $\phit_0(E,\xh)$ in terms of the Fock operator $a(E \ph)$.   In the asymptotic future, $\varphi$ is described by the free-field expression
\be
\varphi(x) \approx  \int \dpp a(\vec{p}) e^{i p \cdot x} +c.c. , 
\ee
where $\dpp \equiv \frac{d^{2m+1}\vec{p}}{(2\pi)^{2m+1} 2 |\vec{p}|}$.
The integral over the sphere $\ph = \vec{p}/|\vec{p}|$ can be evaluated via saddle point in the $r \to \infty$, $u=$const. limit, yielding an expansion of the type (\ref{phiexp}) with
\be
\phi_0(u,\xh) = \frac{e^{-i \pi m/2}}{2 (2 \pi)^m} \int_0^\infty \frac{d E}{2 \pi}E^{m-1} a(E \xh) e^{-i Eu} + c.c.
\ee
We thus obtain,
\be
\phit_0(E,\xh) =   \frac{(-i)^m E^{m-1}}{2 (2 \pi)^m} a(E \xh)  ,  \quad  \text{for} \quad E>0 \label{phia}
\ee
and complex conjugated expression for  $E<0$. Substituting (\ref{phia}) in (\ref{sigmas}) and comparing with (\ref{Qs}) one finds\footnote{Here we wrote the $E \to 0$ limit in (\ref{sigmas}) with the symmetric prescription: $\lim_{E \to 0^+}\frac{1}{2}(\phit_0(E,\xh) + \phit_0(-E,\xh))$.}
\be
\ss(\xh) = \K \Qs(\xh).
\ee 
We finally discuss the hard part of the charges. Substituting (\ref{phia}) in (\ref{sigmah})  yields
\be
\sh(\xh) =  \frac{g}{8(m-1) (2 \pi)^{2m}} \int_{0}^{\infty} \frac{dE}{2 \pi} E^{2m-2} a^\dagger(E \xh) a(E \xh).
\ee
Using (\ref{Kinv}) it is straightforward to verify
\be
\sh(\xh) = \K \Qh(\xh).
\ee 
This concludes the proof of Eq. (\ref{sigQ}). The inverse relation (\ref{Qsig}) follows automatically.

\section{Massive fields} \label{sec4}
In this section we study contribution from massive fields. We thus take  $g' \neq 0$ in (\ref{Lint}). For simplicity we set $g=0$ so that    the field equation for $\varphi$ is   % only contribution to the hard charge will come from the massive field.
\be
\square \varphi =- \frac{g'}{2} \psi^2 . \label{eqm}
\ee
  Our aim is to show that relations (\ref{sigQ}) and (\ref{Qsig}) still hold in this case. The soft part of the charges is the same as before, so we   focus on the hard part of the charges. For $Q(\qh)$ this is given by the last term in (\ref{Qq}),
  \be
 \Qh(\qh) = -\frac{g'}{2} \int \dpp \frac{b^\dagger(\vec{p})  b(\vec{p})}{p \cdot q} , \label{Qhm}
\ee
  whereas for $\sigma(\xh)$ it  is given by (recall  Eq. (\ref{sigma2}))
\be
\sh(\xh) =  \phi_{m-1}(u=\infty,\xh).  \label{shm}
\ee
As  in \cite{ccm} we  proceed by studying the field equations at time-like infinity and then compare the null and time-infinity expansion in their common range of validity. % by relating the $u \to \infty$ limit in (\ref{shm}) to the time-infinity asymptotic value of $\psi^2$ which in turn can be related to $\Qh(\qh)$. 
To this end, we  switch to  hyperbolic coordinates
\be
x^\mu = \t Y^\mu(y), \quad Y^\mu(y) Y_\mu(y) = -1, 
\ee
where $y \equiv y^\alpha, \alpha = 1, \ldots, 2m+1$ parametrize the unit hyperboloid $\H$. In these coordinates the Minkowski metric takes the form
\be
ds^2 = - d \t^2 + \t^2 h_{\alpha \beta} d y^\alpha d y^\beta
\ee
with $h_{\alpha \beta}$ the unit hyperboloid metric. We start by studying the $\t \to \infty, Y^\mu = $const. behavior  of the massive field $\psi$. At late times the field is described by the usual  Fourier expansion
\be
\psi(x)  \approx \int \dpp b(\vec{p}) e^{i p \cdot x}   +c.c. ,
\ee
where $\dpp= \frac{d^{2m+1}\vec{p}}{(2\pi)^{2m+1} 2 \sqrt{\vec{p}^2+\mu^2}}$. In the $\t \to \infty$ limit the momentum integral can be evaluated by a saddle-point yielding\footnote{Up to  phases which  play no role in the analysis.}
\be
\psi( \t Y) =  \frac{\mu^{m-1/2}}{2 (2 \pi \tau)^{m+1/2}} b(\mu \vec{Y}) e^{ -i \mu \t  }  + c.c + \ldots
\ee
 From here we conclude that the leading $\t \to \infty$ asymptotics of the RHS term in  (\ref{eqm}) is given by
\be
- \frac{g'}{2} \psi^2 = \frac{j(y)}{\t^{2m+1}} + \ldots \label{psi2j}
\ee
with
\be
j(y) =-  \frac{g' \mu^{2m-1}}{4 (2 \pi)^{2m+1}}  b^\dagger(\mu \vec{Y}) b(\mu \vec{Y}). \label{defj}
\ee
The field equation (\ref{eqm}) together with (\ref{psi2j})  implies a $\t^{2m-1}$ asymptotic fall-off for $\varphi$
\be
\varphi(\t,y) = \frac{\phih(y)}{\t^{2m-1}} + \ldots \label{phitaufall}
\ee 
with $\phih(y)$ satisfying a Poisson-type equation on $\H$:
\be
(\D^2+(2m-1) ) \phih=j. \label{poisson}
\ee
Eq. (\ref{poisson}) can be solved by Green's functions methods,
\be
\phih(y) = \int d^{2m+1} y'  \sqrt{h} \,  \G(y;y') j(y') \label{solpoiss}
\ee
where the relevant Green's function is
\be
\G(y;y') = - \frac{1}{(2m-1) \vol(S^{2m})} \left[-1 + (Y \cdot Y')^2\right]^{-m+1/2} \label{bulkG}
\ee
with
\be
\vol(S^{2m})= 2^{2m+1}\pi^m \frac{m!}{(2m)!}
\ee
the volume (or area) of the unit $2m-$sphere.

We now use (\ref{solpoiss}) to express $\sh(\xh)$ (\ref{shm}) in terms of $j(y)$. We first choose coordinates $y^\alpha=(\rho,\xh)$  such that
\be
Y^\mu = (\sqrt{1+\rho^2}, \rho \, \xh)
\ee 
with $\rho>0$ and $\xh$ a unit vector parametrizing points on $S^{2m}$. The conformal boundary of $\H$ is obtained by taking $\rho \to \infty$ with fixed $\xh$. In appendix \ref{consistencyapp} it is shown that consistency of the null and time-infinity expansions for $\varphi$ implies
\be
\sh(\xh) = \lim_{\rho \to \infty} \rho^{2m-1} \phih(\rho,\xh). \label{consistency}
\ee
The right hand side of (\ref{consistency}) can be evaluated from expression (\ref{solpoiss}) yielding
\be
\sh(\xh) =  \frac{1}{(2m-1) \vol(S^{2m})} \int d^{2m+1} y \sqrt{h} \, j(y) [Y \cdot (1,\xh)]^{1-2m}, \label{shm2}
\ee 
where  $Y \cdot (1,\xh) \equiv -Y^0 + \vec{Y} \cdot \xh$ is the Minkowski inner product between $Y^\mu$ and the null vector 
$(1,\xh)$. 

We finally wish to compare (\ref{shm2}) with $\Qh(\qh)$.  To this end we first express the momentum integral in (\ref{Qhm}) as an integral over $\H$ by doing the change of variable $\vec{p}= \mu \vec{Y}(y)$, resulting in 
\be
\Qh(\qh) =  \int d^{2m+1} y \sqrt{h} \, \frac{j(y)}{Y \cdot q}. \label{Qhm2}
\ee
The relation  between (\ref{shm2}) and (\ref{Qhm2}) can then  be established in the `inverse' form (\ref{Qsig}),
\be
\Qh(\qh)   = \int_{S^{2m}} d^{2m} \xh \, K^{-1}(\qh,\xh) \sh(\xh) \label{Qshm}
\ee  
by means of a `shadow transform' identity\footnote{The role of the shadow transform in the context of  spin 1 and 2 soft theorems was recently discussed in \cite{kapecmitra}.} \cite{pasterski} which in our conventions reads:
\be
\int d^{2m} \xh \, (1- \qh \cdot \xh)^{-1} [Y \cdot (1,\xh)]^{1-2m}=  \frac{(4 \pi)^{m}}{2} \frac{(m-2)!}{(2m-2)!} \frac{1}{Y \cdot q}.
\ee
The converse relation  $\sh(\xh) =  \K \,  \Qh(\xh)$ then follows automatically. % from (\ref{Qshm})

\section{Smeared charges and spacetime current} \label{sec5}
Given a function $\lambda$ on the sphere 
\be
\lambda : S^{2m} \to \reals
\ee
we define the smeared charge by
\be
\sigma[\lambda] := -2 (m-1) \int d^{2m}\, \xh \lambda(\xh) \sigma(\xh). \label{siglam}
\ee
(In particular, the smearing $\lambda(\xh) = (-1+ \qh \cdot \xh)^{-1} $  yields the charge  $Q(\qh)$.) 
As in \cite{ccm}, we will show that the smeared charge can be written in terms of a  spacetime current $j^a=\partial_b k^{ab}$, 
\be
k^{ab}= \sqrt{\eta}\left( (\nabla^a  \Lambda \varphi - \nabla^a  \varphi  \Lambda)X^b - (a \leftrightarrow b) \right), \label{kab}
\ee
where $X^a\partial_a = x^\mu \partial_\mu$ is the dilatation vector field and $\Lambda$ a space-time scalar determined by $\lambda$ according to:%\footnote{Expression (\ref{Lam}) is a particular case of more general type of fields $\Lambda_n(x) \sim (x\cdot~x)^{n} \int d^{2m} \qh \, \frac{\lambda(\qh)}{(q \cdot x)^{m+n}}$ that satisfy $\square \Lambda_n =0$ \cite{pasterski}. Large gauge parameters of  spin-$s$ fields in harmonic gauge can be obtained by taking $s-1$ derivatives to $\Lambda_n$ with  $n=m+s-1$. See \cite{ccm} for a discussion in four spacetime dimensions.}
\be
\Lambda(x) = c_m (-x \cdot x)^{m-1} \int d^{2m} \qh \, \frac{\lambda(\qh)}{(-q \cdot x)^{2m-1}}. \label{Lam}  %\frac{2}{(4 \pi)^m}\frac{(2m-2)!}{(m-2)!}
\ee
with 
\be
c_m=\frac{2(m-1)}{(2m-1)\vol(S^{2m})}
\ee
a normalization constant.
The field (\ref{Lam}) satisfies the free wave equation
\be
\square \Lambda =0 ,
\ee 
with certain fall-off properties we now describe. 

In the $r \to \infty, u=$const null infinity limit, one can verify (\ref{Lam}) implies  
\be
\Lambda(x) = \sum_{n=1}^{m-1} r^{-n} \Lambda_n(u,\xh) +O(r^{-m})
\ee
with
\be
\Lambda_1(u,\xh) = \lambda(\xh), \quad \Lambda_n(u,\xh) = O(u^{n-1}). \label{lamn}
\ee
On the other hand, in hyperbolic coordinates adapted to time-like infinity one has
\be
\Lambda(\t,y) = \t^{-1} \Lh(y) \label{lamtaufall}
\ee
with
\be
\Lh(y) = c_m \int d^{2m} \qh \, \frac{\lambda(\qh)}{(-q \cdot Y)^{2m-1}}. \label{lhint}
\ee
Expression (\ref{lhint}) can alternatively be interpreted as the solution to the boundary-value problem
\be
(\D^2+(2m-1)) \Lh=0,  \quad \Lh(\rho,\xh) = \rho^{-1}\lambda(\xh) + O(\rho^{-3}) \label{eqlamh}
\ee
(see appendix \ref{applam} for further details). 

We now show that the charge  (\ref{siglam}) can be understood as arising from the spacetime current $j^a=\partial_b k^{ab}$. We follow the same strategy as in \cite{clqed} in which one integrates the current over a $t=$const spatial slice and evaluates the integral in the $t \to \infty$ limit. Two contributions arise in this limit, depending on whether the  limit is taken along null or time-like directions.
%\be
%\sigma[\lambda]= \sigma^{\I}[\lambda]+ \sigma^{\H}[\lambda]
%\ee
The null infinity contribution is given by 
\be
\sigma_{\I}[\lambda] = \lim_{t\to \infty}  \int du d^{2 m} \xh \partial_u k^{ru}  \label{sigmascri}
\ee
where $u=t-r$ and
\be
k^{ru}= r^{2m}\left[( \partial_r \Lambda \varphi  - \partial_u  \Lambda \varphi)u +(\partial_r  \Lambda  \varphi) r\right] - [\varphi \leftrightarrow \Lambda]. \label{kru}
\ee
To evaluate (\ref{sigmascri}) we first substitute the $1/r$ expansions for $\varphi$ and $\Lambda$ in (\ref{kru}). This leads to a $1/r$ expansion for $k^{ru}$ of the form
\be
k^{ru} = r^{m}\left(\sum_{n=1}^{m} r^{-n} k^{ru}_{n} \right) + O(r^{-1}).
\ee
We note that  $k^{ru}$ diverges in the null infinity limit. However the potentially divergent terms in (\ref{sigmascri}) integrate to zero due to the  $u \to \pm \infty$ fall-offs. Indeed, from (\ref{fallu}) and  (\ref{lamn}) one verifies
\be
\begin{array}{ccl}
k^{ru}_n & = & O(|u|^{n-m-\epsilon}) \quad n=1,\dots m-1 \\
k^{ru}_m & = & O(1)+ O(|u|^{-\epsilon})
\end{array}
\ee
and so the would-be divergent terms do not contribute to the charge. The finite contribution comes from 
\be
k^{ru}_m(u,\xh)=2(m-1) \phi_{m-1}(u,\xh) \lambda(\xh) + O(|u|^{-\epsilon})
\ee
and so
\be
\sigma_{\I}[\lambda] =   2(m-1) \int du d^{2 m} \xh   \lambda(\xh) \partial_u  \phi_{m-1}(u,\xh).
\ee
From Eq. (\ref{sigma2}) we see this  corresponds to the null-infinity contribution to (\ref{siglam}).

We now evaluate the contribution from time-infinity.  To this end we switch to hyperbolic coordinates adapted to time-infinity. From the $\tau \to \infty$ fall-offs for $\varphi$ and $\Lambda$ one finds 
\be
\lim_{\t \to \infty} k^{\t \alpha} = \sqrt{h}(\D^{\alpha}\phih \Lh- \D^\alpha \Lh \phih) , 
\ee
 leading to a time-infinity contribution to the charge:
\be
\sigma_{\H}[\lambda]= \lim_{\t \to \infty} \int d^{2m+1}y  \, \partial_\alpha k^{\t \alpha} =  \int d^{2m+1}y \sqrt{h} \, j(y) \Lh(y) \label{Qh3}
\ee
where in the last equality we used the field equations (\ref{poisson}) and (\ref{eqlamh}). 

On the other hand, according to Eq. (\ref{consistency}) the `time-infinity' contribution to (\ref{siglam}) should be given by:
\be
-2 (m-1) \lim_{\rho \to \infty} \rho^{2m-1}  \int d^{2m}\, \xh \lambda(\xh) \ \phih(\rho,\xh) \label{Qh4}
\ee
That (\ref{Qh4}) coincides with (\ref{Qh3}) can be verified by writing $\phih$ in terms  of $j$ according to (\ref{solpoiss}), noting that
\be
\lim_{\rho  \to \infty} \rho^{2m-1} \G(\rho,\xh;y') = -\frac{1}{2(m-1)}  \frac{c_m }{(-q \cdot Y')^{2m-1}},\label{bbbody}
\ee
and  using Eq.  (\ref{lhint}). See appendix \ref{applam} for further details.

%From (\ref{shm2}) and (\ref{lhint}) one readily checks (see appendix \ref{applam}) that expression (\ref{Qh3}) corresponds to the massive `hard' part of (\ref{siglam}). 

To summarize, by evaluating  the spacetime current $\partial_b k^{ab}$ in a $t=$const. slice with $t \to \infty$ one finds two contributions at null and time-like infinity such that their sum gives the smeared charge (\ref{siglam}) 
\be
\sigma[\lambda]=\sigma_{\I}[\lambda]+\sigma_{\H}[\lambda].
\ee

\section{Conclusions}

Following upon the work \cite{ccm}, in this paper we studied the asymptotic charges associated to a  spin zero analogue of Weinberg's soft  theorem in higher even dimensions.  Whereas it is relatively straightforward to recast the soft theorem as a conservation of charges $Q(\qh)$ in the sense of Eq. (\ref{QSSQ}), it is less obvious how to interpret these charges  in terms of asymptotic fields. Here we provided such interpretation,  for the case of both massless and massive `hard' particles. 

The interpretation of $Q(\qh)$ in terms of asymptotic fields allows one to obtain a spacetime expression for smeared charges. % that contain the same information of the charges $Q(\qh)$. 
 As in  gravity and electrodynamics,  these smeared charges can be written as integrals of total-derivative currents, $j^a= \partial_b k^{ab}$. But unlike those cases, there is so far no symmetry understanding for the tensor $k^{ab}$: In gravity and electrodynamics $k^{ab}$ are Noether forms of local symmetries \cite{avery}; but there are no local  symmetries associated to scalars! The question remains as to what is the  appropriate symmetry interpretation for the `soft scalar' asymptotic charges.
\\

%There is a growing list of examples where soft factorization theorems can be understood in terms of conserved asymptotic charges \cite{lecstrom}. The simplest and most studied examples are those

%The firstly studied cases of (leading) soft theorems in gravity and electrodynamics Following \cite{mmc}, in this paper we have enlarged the list by including the case of massless $\varphi^3$ theory in higher even dimensions (or the case of massive particles 

%Given the recent understanding of soft theorems in terms of asymptotic charges, one may wonder how far this idea can be pushed 

\noindent {\bf Acknowledgements }\\
We would like to thank Alok Laddha and Sebastian Mizera for many  key discussions. This work was supported by Pedeciba, Fondo Clemente Estable FCE 1 2014 1 103803,  and Perimeter Institute for Theoretical Physics. Research at Perimeter Institute is supported by the Government of Canada through the Department of Innovation, Science and Economic Development and by the Province of Ontario through
the Ministry of Research and Innovation.

\appendix
\section{Operator $\K$ and Eq. (\ref{Kinv})} \label{appK}
The operator $\K$ (\ref{K}) is proportional to the composition of the $(m-1)$  differential operators:
\be
\Delta_n = D^2 - \mu_n; \quad \mu_n =(m-n)(m+n-1), \quad  n = 1, \ldots, m-1, \label{deln2}
\ee
where $D^2$ is the Laplacian on $S^{2m}$ and $m \geq 2$.  It can be easily checked that $\mu_n >0$ in the range (\ref{deln2}). Since the eigenvalues of the Laplacian $D^2$ are non-positive, it then follows that zero is not an eigenvalue of  $\Delta_n$. Thus  $\Delta_n$ is invertible and so is $\K$.

To find the inverse of $\K$ we start by inverting $\Delta_1$. Its Green's function $G_1(\qh,\xh)$,
\be
\Delta_1 G_1(\qh,\xh) = \delta^{(2m)}(\qh,\xh) \label{Del1G1}
\ee
is found to take the simple form\footnote{The reason for the simplicity of $G_1$ is that $\Delta_1$ is conformally equivalent to the flat Laplacian on $\reals^{d}$:  $\Delta_1= D^2 - \frac{d-2}{4(d-1)}R$ with $R=d(d-1)$ the scalar curvature on $S^d$. In terms of $\Delta_1$, the operators (\ref{deln2}) take the simple  form: $\Delta_n=\Delta_1 + n(n-1)$.} 
\be
 G_1(\qh,\xh) = \frac{-1}{2^m(m-1) \vol(S^{2m-1})}(1- \qh \cdot \xh)^{1-m} \label{G1}
\ee
where 
\be
\vol(S^{2m-1})= \frac{2 \pi^m}{(m-1)!}
\ee
is the volume of the unit $S^{2m-1}$ sphere. We now discuss the $m=2$ and $m>2$ cases separately. For $m=2$, $\K= -(4 \pi)^{-2} \Delta_1$ and so $K^{-1} = -(4 \pi)^2 G_1 = 2 (1- \qh \cdot \xh)^{-1}$ which coincides with  expression  (\ref{Kinv}). For $m>2$ we use the identity \cite{mirbabayi},
\be
\Delta_{m-p} (1- \qh \cdot \xh)^{-p} = 2 p (p+1-m) (1- \qh \cdot \xh)^{-(p+1)}
\ee
%(which can be established by direct computation)
 from which it follows that
\be
\prod_{n=2}^{m-1} \Delta_n (1-\qh \cdot \xh)^{-1} = (-1)^m 2^{m-2}(m-2)!^2 (1-\qh \cdot \xh)^{1-m} .\label{prod2}
\ee
Finally, acting with $\Delta_1$ on (\ref{prod2}) and using (\ref{Del1G1}),  (\ref{G1}) one arrives at the desired result
\be
\K (1-\qh \cdot \xh)^{-1}  =  \frac{1}{2(m-1)} \delta^{(2m)}(\qh,\xh).
\ee

\section{Eq. (\ref{consistency})} \label{consistencyapp}
The asymptotic properties of $\varphi$ near null infinity described in section \ref{scrisec} imply certain asymptotic properties at time-infinity which we now explore. 

The idea is to consider a double $r \to \infty$ \emph{and} $u \to \infty$ expansion of $\varphi$. The $r \to \infty$ expansion is the one given in (\ref{phiexp}) which we take to hold for all $n$:
\be
\varphi(r,u,\xh)= r^{-m}\sum_{n=0}^{\infty} r^{-n}\phi_n(u,\xh) . \label{phiexp2}
\ee
Each $\phi_n(u,\xh)$ has a $u \to \infty$ behavior as in  (\ref{fallu}),
\be
\phi_n(u,\xh) =   \phi^+_n(\xh) u^{n-(m-1)-\epsilon}  + \delta_{n  m-1} \sh(\xh) \quad \text{for }  u \to \infty
\ee
where we included the   additional $O(1)$ term that appears when  $n=m-1$ (identified as the `hard' charge density due to massive particles, see Eq. (\ref{shm})). 
The leading $r \to \infty, u \to \infty$ behaviour of $\varphi$ is thus given by
\be
\varphi = \sum_{n=0}^{\infty} r^{-(m+n)} u^{n-(m-1)-\epsilon} \phi^+_n(\xh) + r^{-(2m-1)}\sh(\xh) + \ldots. \label{phru}
\ee
We now wish to identify this expansion with a $\t \to \infty, \rho \to \infty$ expansion at time-infinity. The relation between the two sets of coordinates is:
\be
r= \rho \t, \quad \quad u = \t (\sqrt{1+\rho^2} - \rho) = \frac{\t}{2 \rho} +O(\rho^{-3})
\ee
Substituting in (\ref{phru}) and keeping only leading terms in $\t \to\infty, \rho \to \infty$, one finds
\be
\varphi= \frac{1}{\t^{2m-1} \rho^{2m-1}} \sh(\xh) + O(\frac{1}{\t^{2m-1+ \epsilon}}) + \ldots  \label{phirhotau}
\ee
Recalling that the $\t \to \infty, \rho=$const. behavior of $\varphi$ is given by
\be
 \varphi(\t,\rho,\xh) = \frac{\phih(\rho,\xh)}{\t^{2m-1}} +O(\frac{1}{\t^{2m-1+ \epsilon}})
 \ee 
we conclude from (\ref{phirhotau}) that 
\be
\lim_{\rho \to \infty} \rho^{2m-1}\phih(\rho,\xh)= \sh(\xh).
\ee

\section{Green's functions at time-infinity} \label{applam}
In this appendix we collect some expressions from \cite{solo,pasterski} in a way geared to our setup. % We start with  general expressions from  \cite{solo,pasterski} and then restrict attention to the case of interest in this paper. 
We will be slightly more general than in the rest of the paper and allow for arbitrary sphere dimension $d$ and arbitrary `conformal dimension' $h$.  The case of interest in the paper is $h=1$. (More generally the value $h=1-s$ is relevant for spin-$s$ leading soft theorem charges; see \cite{ccm} for a discussion in four spacetime dimensions.)

We start by constructing a spacetime function $\Lambda(x)$ out  of a function on the sphere $\lambda: S^d \to \reals$ and a conformal dimension $h \in \reals$ by \cite{solo,pasterski}:
\be
\Lambda(x) := \frac{2^h}{ (4 \pi)^{d/2}}\frac{\Gamma(d-h)}{\Gamma(\frac{d}{2}-h)} (-x \cdot x)^{\frac{d}{2}-h} \int d^d \qh \frac{\lambda(\qh)}{(-q \cdot x)^{d-h}}.
\ee
One can verify $\Lambda(x)$ satisfies the free wave equation
\be
\square \Lambda=0 \label{boxlam}
\ee
as well as the scaling property $t^h \Lambda(t x)= \Lambda(x)$. The role of the overall normalization constant becomes clear below. 

In terms of hyperbolic coordinates adapted to time-like infinity (see section \ref{sec4}), the function takes the form
\be
\Lambda(\t,y) = \t^{-h} \Lh(y) \label{lamh}
\ee
where
\be
\Lh(y) = \int d^d \qh \, G(y;\qh) \lambda(\qh) \label{lamhh}
\ee
with
\be
G(y;\qh)=  \frac{2^h}{ (4 \pi)^{d/2}}\frac{\Gamma(d-h)}{\Gamma(\frac{d}{2}-h)} (-q \cdot Y)^{h-d}. \label{greenh}
\ee
From Eqns. (\ref{boxlam}) and (\ref{lamh}) it follows that $\Lh$ satisfies 
\be
(\D^2+h(d-h)) \Lh =0 \label{elliptic}
\ee
 on $\H$. On the other hand, Eq. (\ref{lamhh}) implies \cite{solo,pasterski}
\be
\lim_{\rho \to \infty} \rho^h \Lh(\rho,\xh)=  \lambda(\qh) .\label{bdylam}
\ee
In other words, $G(y;\qh)$  is the Green's function for the boundary value problem given  by Eqns. (\ref{elliptic}), (\ref{bdylam}).

At time-infinity there are also fields $\phih$ that capture the contribution to charges due to massive particles \cite{clqed,clmgrav,eyhe,ccm}. These fields are in a sense  `dual' to $\Lh$  
and satisfy Poisson-type equations 
\be
(\D^2+h(d-h) ) \phih=j ; \quad \quad \phih(\rho,\xh) \stackrel{\rho \to \infty}{=}O(1/\rho^{d-h}), \label{elliphih}
\ee
where $j$ is a source term that is quadratic in the massive particle Fock operators. 

If $\G(y;y')$ is the Green's function for the elliptic problem (\ref{elliphih}) then its $\rho \to \infty$ asymptotic value is related to the Green's function (\ref{greenh}) by \cite{penedones}:
\be
 \lim_{\rho \to \infty} \rho^{d-h} \G(\rho,\xh; y') = -\frac{1}{(d-2h)} G(y';\xh), \label{bulkbdyGs}
\ee
 a well-known fact in the AdS/CFT literature. 
 
For the case of interest in the  present paper we  set 
%\footnote{More generally the value $h=1-s$ is relevant for spin-$s$ (leading) soft theorem charges.} 
$h=1 $ and $d=2m$. % (after taking $s-1$ derivatives of $\Lambda$ so as to get correct tensorial index structure; See \cite{ccm} for a discussion in four spacetime dimensions).
 In this case $\G$ is given by Eq.  (\ref{bulkG}), $G$ can be read-off from Eq. (\ref{lhint}) (or alternatively from Eq. (\ref{greenh})), and relation (\ref{bulkbdyGs}) becomes 
\be
 \lim_{\rho \to \infty} \rho^{2m-1} \G(\rho,\xh; y') = -\frac{1}{2(m-1)} G(y';\xh) ,  \label{bulkbdyGs2}
\ee
which  corresponds to Eq. (\ref{bbbody}) in the  body of the paper.

%\section{Charges at spatial infinity}

%\subsection{Eq. (\ref{})}

\end{document}